# Overt and covert paths for sound in the auditory system of mammals,

B. Auriol[1*], B. Auriol[2], J. Béard[3], B. Bibé[4], J.-M. Broto[5*], D.F. Descouens[6], L.J.S. Durand[7], J.-P. Florens[8], F. Garcia[9], C. Gillieaux[10], E.G. Joiner[11], B. Libes[12], P. Pergent[13], R. Ruiz[14], C. Thalamas[15].

## ABSTRACT

Current scientific consensus holds that sound is transmitted, solely mechanically, from the tympanum to the cochlea via ossicles.

However this theory does not explain the hearing extreme quality regarding high frequencies in mammals. So, we propose a bioelectronic pathway (the covert path) that is complementary to the overt path.

We demonstrate experimentally that the tympanum produces **piezoelectric** potentials isochronous to acoustic vibrations thanks to its collagen fibers and that their amplitude increases along with the frequency and level of the vibrations. This finding supports the existence of an electrical pathway, specialized in transmitting high-frequency sounds that works in unison with the mechanical pathway. A bio-organic triode, similar to a field effect transistor, is the key mechanism of our hypothesized pathway. We present evidence that any deficiency along this pathway produces hearing impairment. By augmenting the classical theory of sound transmission, our discovery offers new perspectives for research into both normal and pathological audition and may contribute to an understanding of genetic and physiological problems of hearing.

## 1. Introduction

The scientific literature of sound transmission and perception is founded on the travelling wave (TW) principle proposed by von Bekesy (Nobel, 1961). The eardrum vibrates in response to sound. These vibrations travel via the three ossicles through the oval window and into the fluid-filled cochlea. Inside the cochlea, acoustic signals are broken down into their component frequencies by the mechanical properties of the basilar membrane, to which the hair cells are attached.

This paper questions the adequacy of current theory to explain the transmission of high frequency sounds (above 3 kHz according to Quix[1]) along the tympano-cochlear pathway. Beginning with the tympanum, we observe that radial fibres of collagen extend from its periphery to its centre.

**1**- MD, Toulouse, France (**\***auriol@free.fr). **2**- Théâtre du Capitole, Toulouse. **3**- LNCMI, CNRS, EMFL, INSA, UGA, UPS, Toulouse. **4**- Emer. Dir. of Research at INRA, Toulouse.  **5**- University Professor, UPS, Toulouse 3, LNCMI CNRS (\*jean-marc.broto@univ-tlse3.fr). **6**- MD (ENT), UMR 5288, CNRS, Toulouse. **7**-  CEMES/CNRS, Toulouse. **8**- TSE, Toulouse. **9**- INRA UR875 Unité de Mathématiques et Informatique Appliquées, Toulouse. **10**- DVM - Montel Veterinary Clinic, Tournefeuille, France. **11**- Department of Languages, Literatures, and Cultures, The University of South Carolina, USA. **12**- MD, ENT, CRA, CHU Purpan, Toulouse. **13**- Veterinary Doctor, Puylaurens, France. **14**- LARA-SEPPIA, Toulouse. **15**- CIC 1436, Université Toulouse-Purpan, INSERM.

Thus the eardrum is a membrane attached to its periphery. At high frequencies, the Chladni model implies local resonances, which fragment its surface into vibrating zones all the more complex as the frequency increases[2].

This implies[3] that ossicular transmission is extremely weak for frequencies above 2 kHz. The magnitude of the transduction decrease particularly in the 2-5 kHz range [4-5] (Cf. fig. 1a and 1b below) and this apparent flaw in current theory has not been convincingly explained[6].

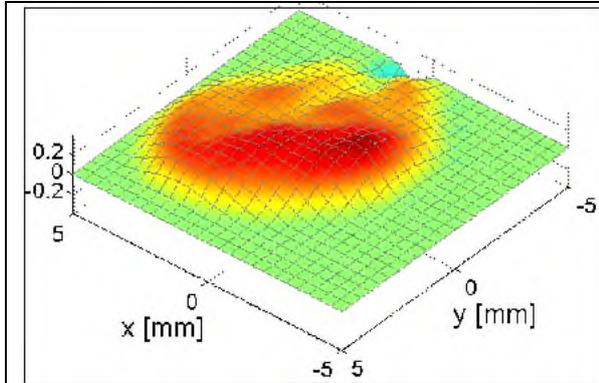

Fig. 1a
frequency 1062 Hz
jbio_201200186_sm_video04.avi

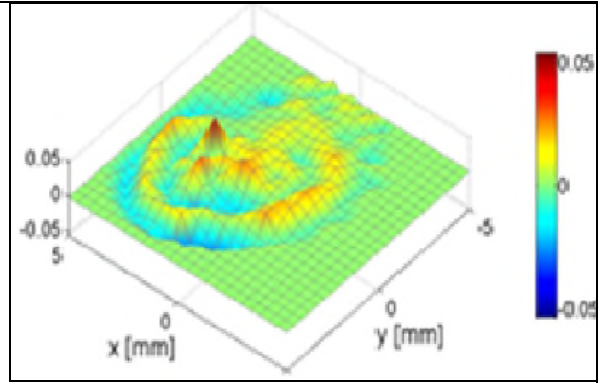

Fig. 1b
frequency 5175 Hz
jbio_201200186_sm_video06.avi

Excerpted instant photographs of OCT videos [from Burkhardt A. et al., Investigation of the human tympanic membrane oscillation ex vivo by Doppler OCT, J Biophotonics,7:434-441 (2012).].
These videos show that the tympanic surface responds to the high frequencies in a much more fragmented way (*Chladni phenomenon*) than to the frequencies under 1500 Hz.

Click here while simultaneously using the Ctrl key : This figure shows the effect of Bessel functions resulting from vibrations imposed on a roughly circular thin membrane. This phenomenon is particularly at work in the vibratory phenomena discovered experimentally by *Chladni*.

The Caminos et al., following figures[7], clearly illustrate the decay of the amplitude transmitted by the eardrum from 1 kHz up to highest frequencies, via the ossicles and up to the stapes.

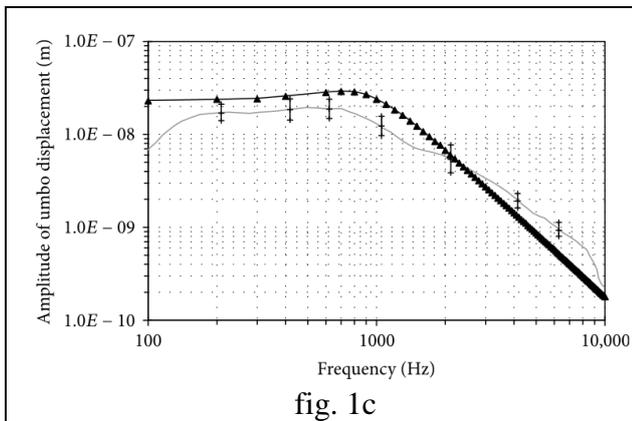

fig. 1c

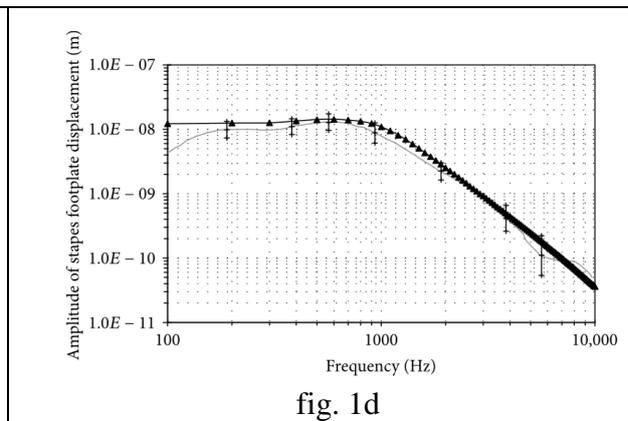

fig. 1d

--Exp. Hato et al. [intact cochlea] (2003)
--▲-- Finite Elements Model (FEM)

This work shows that the original TW is insufficient for the processing of high and medium frequencies.

According to Nakajima et al[8] "The significant sound pressures measured at certain frequencies (e.g. 6 kHz) after ossicular interruption suggest that sound is transmitted to both [cochlear] scalae through a path independent of the ossicular chain". Furthermore, cetaceans and other sea mammals develop only vestigial parts of the external and middle ear yet have extreme hearing capabilities[9] (Cf. Si 08A). It seems, then for Röösli et al[10], that "a mechanism independent of the chain of ossicles is necessary for optimal transmission of high frequency sounds".

Our search for a mechanism that would overcome this low-pass effect has led us to hypothesize a sound signal path originating from the piezoelectricity of the tympanum and bone collagens.

Furthermore, contrarily to a common supposition, the electric response of eardrum is not uniquely due to the "backward electrical waves" coming from cochlea OHCs (Cf. Si 05).

Examining the Deiters cells and the outer hair cells (OHC) at the other end of the tympano-cochlear pathway[11], we note that stereocilia crowning the OHC have a mechano-electrical activity that transduces the acoustic TW into electrical signals[12] (Cf. Si 10). This transduction of acoustic waves, and the transmembrane transmission of resultant electrical signals, should involve the RC time constant of the plasma membrane and ion channels, resulting logically, in low-pass filtering (<1 kHz)[13]. However, some mammals hear at frequencies above 200 kHz[14]. At these high frequencies, the relaxation time constant would be $\tau \leq 1/(2\pi \times 200 \text{ kHz}) \leq 1$ µs, i.e. an order of magnitude faster than that found for ion channels[3]. Further, there is very intense debate[15] about currently accepted concepts. Several models have been advanced, but none has yet been experimentally verified, and the invoked mechanisms could not allow the transmission of frequencies higher than 12 kHz[16].

Thus, we have seen that, at the beginning of the tympano-cochlear pathway, as at its end, the high frequencies should be weakened.

## 2. The tympanum, a piezo-electric bio-electret

The triple-helical collagen molecules are organized hierarchically into fibrils, fibers, and bundles. Sounds produce piezoelectric potentials due to the collagen fibers[17] of the tympanum. Fibers, like fibrils, are piezoelectric bioelectrets[18], having a negative pole (C-terminal) and a positive pole (N-terminal).

The voltage of the piezoelectric potentials due to isolated fibers[19] is much higher (up to ten millivolts) than the potentials we measure on bundles of millimetric fibers.

Furthermore ear canal obstruction, physical separation between eardrum and the cochlea or general anesthesia (ketamine) confirm that electrical potentials, isochronous to acoustic stimuli, do exist at the local level of collagen structures and are measurable independently of the activity of OHCs. So there is not any ambiguity at all between these two electrical activities.

Could the recorded potentials be the simple result of artifacts dependent on ambient electromagnetic phenomena such as microphonic potentials generated by coaxial conductors, or the speaker? (Cf. Si 04).

If this were the case, the observed potentials should persist even in the absence of collagen fiber structures. In fact, during measurements concerning the patellar tendon we found that its replacement with a metal prosthesis (non collagenic) abolished the electrical response of the considered knee, while the contra lateral knee, which had retained its tendon and was not equipped of prosthesis, responded electrically as the knees of all other subjects of the group.

We verified that, in the absence of an acoustical signal, the measured voltage was zero, whereas for almost all the observed series, non-zero microvoltages could be measured.

---

[3] Yet "When the OHC electrical frequency characteristics are too high or too low, the OHCs do not exert force with the correct phase to the OC mechanics so that they cannot amplify. We conclude that the components of OHC forward and reverse transduction are crucial for setting the phase relations needed for amplification" (Nam J-H, Fettiplace R (2012) Optimal Electrical Properties of Outer Hair Cells Ensure Cochlear Amplification. PLoS ONE 7(11): e50572. https://doi.org/10.1371/journal.pone.0050572).

# 3. Material and Methods

### 3.1. Aim of the study

This study was designed to test our hypothesis that the collagen fibers of the tympanum are piezoelectric. Our protocol involved stimulating the tympanum at various frequencies and using various pressures. We measured the potentials resulting from this stimulation to determine if their electric properties were dependent upon the amplitude (dB SPL) and the acoustic frequency (Hz).

### 3.2.Types of Collagen

Collagens II and I are piezoelectric histological components of eardrum. Their properties are very similar and we made measurements not only on collagen II of eardrum, but also on collagen I of the patellar tendon with the purpose to know at best their piezoelectric properties (Cf. Si 01C; Si 01D). It is possible to detect an electric potential isochronous to the acoustic vibration between an indeterminate point of the tympanum and the mastoid bone[20]. It does not follow necessarily, however, that the potential measured in this type of experiment is produced by the Outer Hair Cells (OHCs). Our methodology[21] allows us to demonstrate, in vivo, and under normal physiological conditions the piezoelectricity both of collagen I in tendons and of collagen II in eardrums.

### 3.3. Measurements tools for eardrum, tendons and bones fibers

For every measurements (eardrums, tendons and bone collagen) we use a lock-in amplifier to drive a loudspeaker. In this manner, we broadcast a sinusoidal sound at about one meter from the target (external auditory conduit, etc.) (Cf. Si 01E, Si 01F, Si 01G ). We position a probe consisting of two electrodes at the center and the periphery of the tympanum, or at the ends of any targeted fibers. This probe captures the piezoelectric response of the fibers when they vibrate in response to the sound sent to the target. The lock-in amplifier makes it possible to select only those electrical responses isochronous to the acoustic stimulation. We measure electrical responses to stimulations at different acoustic frequency levels.

A Lock In Amplifier, via an electric wave of $V_{hp}$ tension (output) creates an acoustic wave at various frequencies diffused by a loudspeaker.

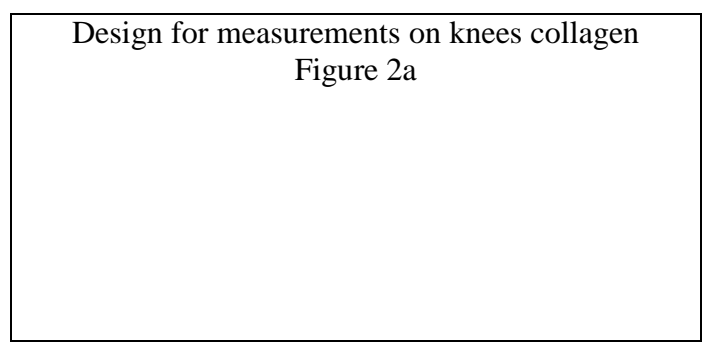

Design for measurements on knees collagen
Figure 2a

### 3.4. Approval for the study

The official title of this research is " Non-invasive Study in vivo of the generation of microvoltages by the tendons and the tympanum when they are subjected to moderate sound stimulations".
This work was promoted by CNRS-INSB (ID RCB N°2012-A01375-38; protocole12 008), Paris (France). This study was carried out in accordance with the recommendations of Institut National des Sciences Biologiques - CNRS PARIS. This protocol was approved by the French Institutional Review Board (IRB/IEC) ad hoc : CPP SOOM 4 N° ID RCB 2012-A01375-38. It was conducted in accordance with "*Good Clinical Practices*" and with French legislation on clinical trials (Loi de santé publique n° 2004-806 du 9 août 2004, Titre V, Chapitre II (recherches biomédicales).

Approval of the study was obtained from the following:
Committee of Experts and National Institute of Physics of the CNRS (Research Laboratory: UPR 8011).
Centre d'Elaboration de Matériaux et d'Etudes Structurales (CEMES / CNRS–Toulouse).
Comité de Protection des personnes (CPP SOOM4, Jan-Apr 2013; N° CPP13006a),
      Agence Nationale de Sécurité des Médicaments (ANSM, B130246-81; March 2013),
      Agence Française de Sécurité Sanitaire des produits de Santé (ASSFAPS, June, 2013).

## 3.5. The Subjects

The subjects (N=35) were healthy and their hearing was "normal", as determined by an ENT.
We did, however, include one subject (D31) who had a cicatricle eardrum and another who needed to use a hearing aid (D33). In the latter case, the subject removed the prosthesis during testing. Other measurements were made on knees and other tendons (fig.2a). After being informed and signing the Inform Consent Form, all subjects gave written informed consent in accordance with the Declaration of Helsinki. Then they were tested individually in a room designated for that purpose. The piezoelectricity of collagen fibers and the impedance of the epidermis and of the dermis vary with temperature. For this reason, the testing room was heated to 22° C in colder weather.

## 3.6. The measurements team

The measurements team consisted of two physicists and three physicians. Two of the physicians were ENTs.
We checked the health and medical history of each subject through a personal interview and a clinical examination (audiogram, otoacoustic emissions) performed by an ENT physician.
Then, one of the ENTs and another physician explained the protocol and installed the subject. Earlier, the two physicists had verified the correct operation of the lock-in amplifier and the connection of the probe conductors to the input of the lock-in.
The physicists set up the left and right loudspeakers (1 m from the subject's ear, and at the same height as his or her ears). They connected the lock-in output to the input of the loudspeaker to be used.
One of the two physicists was responsible for regulating the device for each of the frequencies to be studied. This included setting the frequency, sensitivity, and the amplitude of the stimulation addressed to the loudspeaker.
The other physicist supervised all these actions and noted the results, i.e., value of the voltage detected by the probe in response to each stimulation in turn.
After each testing session, the team met with the subject to obtain reactions and to determine the degree of "discomfort," if any, felt by the subject for either ear. The team also met to compare observations they had made during the testing.

## 3.7. Lock-In Amplifier

To demonstrate, in vivo, and under normal physiological conditions, the piezoelectricity of tympanum fibers, we used either a digital lock-in amplifier ("Stanford SR830") or a dual-phase analog lock-in amplifier ("EG and G, M 5210") to drive a loudspeaker.
A lock-in amplifier can extract a signal with a known carrier wave from an extremely noisy environment. Signals up to $10^6$ times smaller than noise components can still be reliably detected.

## 3.8. Loudspeakers

In our protocol, the lock in amplifier, via an electric wave of $V_{hp}$ tension (output), created an acoustic wave at selected frequencies broadcast by a loudspeaker. The loudspeaker was either a [Harman/Kardon: DP/N 0865DV][4] or a [Yamaha HS 50]; For frequencies higher than 20 kHz, we used a [Conrad TE300 tweeter].

### 3.9. Tympanic Probe

Furthermore, the conformation of the eardrum is quite variable depending on the individual, and the experimenters must adapt to it. For example, according to some of them, measures were taken under better conditions when one of the two electrodes was resting near the handle of the hammer but not on it, while the other electrode was in place near the annulus tympanicus, but staying on the periphery of the tympanic membrane without encroaching on the annulus tympanicus (i.e. local recordings).

The advantage of that differential recording technique[22] was that it allowed us to determine the source of the potential[23].

Thus we tried two configurations (Fig. 2b):
(a) two electrodes equipped with a manipulable rigid elbow and flexible electrodes, easy to handle, easily affixed to appropriate parts of the eardrum, but not allowing standardization of the inter-electrodes distance... Because of the obliquity of the eardrum, other experimenters might wish to use a probe with electrodes that are easy to mold extemporaneously.
(b) two electrodes attached to a plastic bracket, ensuring a constant distance, but more difficult to handle.

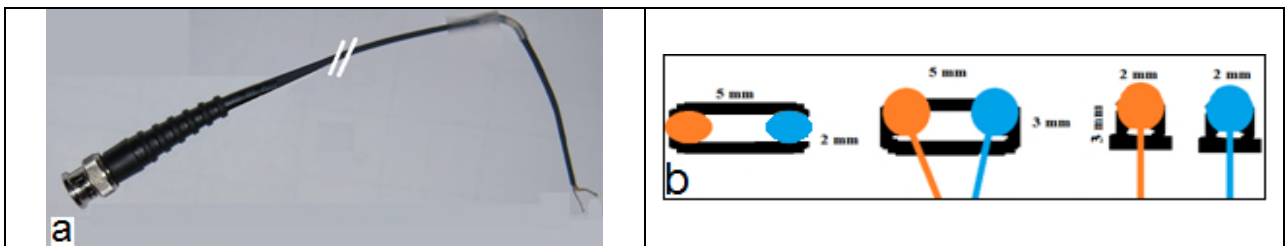

Figure 2b
Probes used for measurement on the eardrums
In the panel [b] we display three possible configurations of proximal contacts couples of the electrodes. The blue and red colors are useful for replication of the experiences between the two ears and between two people ears as well.

The shielding braid is put in contact with the peripheral structure of the tympanum, away from the handle of the malleus (Fig. 3). The central copper wire is put in contact with the central structure of the tympanum: either umbo (direction "A") or handle of the malleus ("G").

---

[4] no longer available

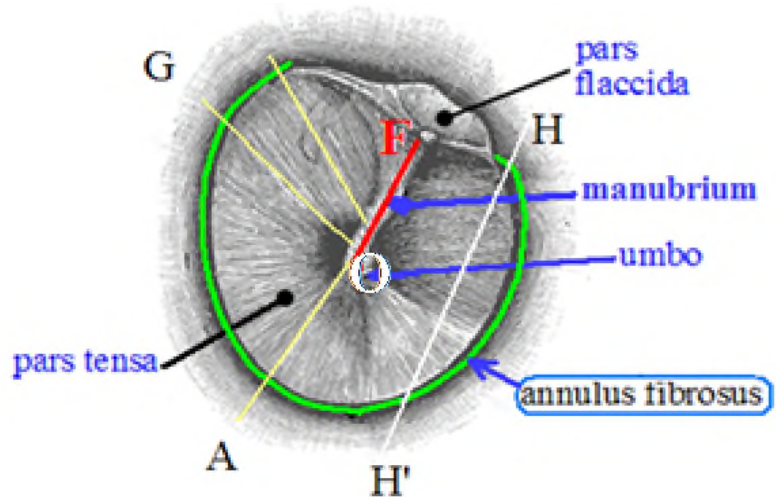

Figure 3
Right Eardrum (photography)
The letters A, F, G, H, H' denote the direction and location of the electrode pairs of the measurement probe for a diversified exploration of the piezoelectricity of the tympanic collagen fibers.

It is possible to detect an electric potential isochronous to the acoustic vibration between an indeterminate point of the eardrum and the mastoid bone[5]. However, it does not necessarily follow that this potential be a microphonic produced by Outer Hair Cells OHCs. On the contrary, our methodology allows us to demonstrate, in vivo, and under normal physiological conditions, that it is the result of the piezo-electricity of the collagen II of the eardrums.

The closer together the electrodes are placed, the smaller the area recorded will be (i.e. local recordings). This is one advantage of the differential recording technique[24]. This "differential technique" is necessary for determining the source of a potential[25].
In order to evaluate the electrical behavior of points belonging to the central structure (manubrium) during acoustic stimulations, electrodes can be placed at two points on the same side of the manubrium (symbolized by an F letter). This system can detect if there is electrical isochronism between these points. On the contrary, electrodes might be placed facing each other on either side of the manubrium (E). This latter system would allow us to capture the activity of a bundle of circular fibers (not completed).
The following letters were added by us: A or G: "radii" types of fibers of collagen ; HH': arbitrary cord joining two peripheral points ; F: manubrium of the malleus ; GAH'H: ie annulus tympanicus (quasi-circle, green color, in the figure). In order to verify the piezoelectric activity of the tympanum, an electrode is placed on the manubrium, another on the periphery according to the straight lines either A or G.
With humans the synchronous electrical responses between two points on the same side (inner or outer) of the annulus tympanicus (HH') are generally impossible to measure.

For measurements on the tympanum (Cf. Si 01G), we used a probe consisting of a coaxial copper cable (1 mm outside diameter). The advantage of coaxial design is that electric and magnetic fields are restricted to the dielectric, with little leakage outside the shield. Electric and magnetic fields outside the cable are entirely prevented from interfering with signals inside the cable. This property makes coaxial cable a good choice for carrying weak signals without interference.
In order not to damage the epidermis of the eardrum, the ends of the two strands were either coated by a small amount of an eutectic solder or shaped to form a smooth closed loop..
The probe consists of two electrodes based on a coaxial cable:

Central copper wire wrapped with a sheet of plastic isolator (with the end of the copper wire used for contacting the eardrum)

Surrounding conductive braid wrapped with a sheet of plastic isolator (with its end used for contacting the eardrum).

For measurements, one electrode was apposed to the center of the eardrum (a point on the posterior side of the manubrium) and the other at its periphery (a point on the posterior limbus). The probe captures the piezoelectric response of the radial tympanic fibers when they vibrate in response to the sound sent to the tympanum. We measured electrical responses to stimulations, at different acoustic and frequency levels. We selected 16 frequencies (125 Hz to 30 kHz) and five sound pressure levels (55 to 80 dB) for this study. All measures were taken on both sides.

An electric wave of $V_{hp}$ tension is sent by the lock-in to the loudspeaker, creating an acoustic wave at the chosen frequency.

Then, the electric tympanic probe receives the piezoelectric response. The voltage of the response is displayed on the digital screen of the lock-in.

## 3.10. Otoscopy for optimal visual access

We used either a Zeiss OPMI 99 microscope (19X magnification) or a KAPS Som62 MAT005 halogen microscope (16X magnification).

## 3.11. Units of measurement

As units of measurement, we used the International System. For acoustic frequencies, we used Hz (cycles/second), and for amplitudes the decibel sound pressure level (dB SPL).

## 3.12. Possible drawbacks with respect to measurements

Our measurements on the eardrum encompass broadcasting a sinusoidal sound at one meter from the external auditory conduit. We were very cautious, yet the measurement of sound amplitude reaching the eardrum from a loudspeaker in a free field may be problematic.

Spatial fluctuations of the sound pressure might be an important factor: a frontal or lateral shifting of a few centimeters between the sound source and the ear of the subject may have significant effects on perceived and measured amplitudes. In despite of all our efforts to assure a strong stability during measurement process, the experimenters could not ensure that the clinician, or his hands, were, or were not, at times, an obstacle dampening the amplitude for the frequencies emitted by the loudspeaker.

We recall moreover, that, in addition to the implied physical parameters, in live measurements, there are psycho-physiological interactions as well. When enabled, the tensor tympani muscle pulls the malleus medially, tensing the eardrum and damping its vibrations (Cf. Si 03A). This is also the case of the stapes muscle and of the smooth muscles that are inserted into the annulus tympanicus. Cerebral commands of OHC activity likewise follow this pattern.

In addition, the position of the electrodes should make it possible to measure the variations of potential in response to the sounds between the two ends of one and the same fiber; This is obviously very difficult to achieve, especially in vivo. It is obvious that our measures are underestimated (Cf. Si 02). Similarly, it is very difficult to standardize the pressure of the electrodes on the skin (Cf. Si 03B).

## 3.13. Anesthesia during measurement on the eardrum

Measurements on the eardrum of humans were made after applying a light anesthesia to the eardrum (*cream EMLA 5%*:  Lidocaine 25 ‰, Prilocaine 25 ‰) this cream was removed afterwards, using vacuum aspiration or wiping with gauze.

We did not use a conductive paste because it would be inappropriate to mix it with the anesthetic paste. Also, the quantity and surface of contact would be very difficult to standardize.

## 3.14. Evaluation of the possibility of pain

Yet, an index (Likert scale, 0 to 5) of "*felt pain*" was recorded.

## 3.15. Differentiation of piezo-tympanic potential and cochlear microphonics

We have considered and evaluated possible technical artifacts [links to ground  Si04A & B, links to epidermis Si04C ;  microphonics of measurement cables  Si 04D].
We have shown that our measurements correspond to a tympanic generation and that it would be a mistake to equate them with cochlear microphonics (Cf. Si 05). This supports our hypothesis that the electrical response of the system cannot be reduced to cochlear microphonic activity (Cf. Si 06).
We performed an experiment to check whether the sound generated a piezoelectric response at the mastoid level, including by blocking the airway.

## 3.16. Statistics

For explaining the potential difference V, we estimated a regression model, log-quadratic in the frequencies F and linear in the sound level L, including a fixed effect component (for each individual and each side).
The model was estimated by ordinary least squares completed by the usual tests (global tests of significance and tests of significance of the parameters) with the R software[26] (*lm* command).
Further analysis of individual characteristics and experimental conditions is presented in Si 14, Si 14A, Si 14B, Si 15, Si 16.

## 4. Results

We tested 35 individuals, aged 17 to 87, for 16 frequencies (125 Hz to 30 kHz) and for five sound pressure levels (55 to 80 dB). All measures were taken on both sides. Due to missing data, the sample size was 453 (Cf. above Material and Methods). This sample constitutes an unbalanced panel data sample estimated by ordinary least squares with the introduction of a so-called "fixed effect" dependent on the individual and the side (left or right). For the raw data, Cf. 10.6084/m9.figshare.5671807.

The empirical evidence of a significant effect of frequency and sound pressure on the difference in potential was verified by statistical analysis using the "R" software. The explained variable is the micro voltage $V_{ijks}$ measured for individual $i$, frequency $j$, sound pressure level $k$ and side $s$ (left or right).

The relation is assumed to be log-quadratic in the frequencies $F$ and linear in the sound level $L$.  The least squares estimation is:

$$\log(1+V_{ijks}) = \alpha_{is} + 0.036616\ L_{iks} - 2.142166 \log(F_{ijs}) + 0.144777\ (\log(F_{ijs}))^2\ + U_{ijks}$$
$$\text{(s.d. = 0.006437) (s.d. = 0.481262) (s.d. = 0.029839)}$$
$$\text{where } U_{ijks} \text{ is a zero mean random residual (Cf. fig 4).}$$

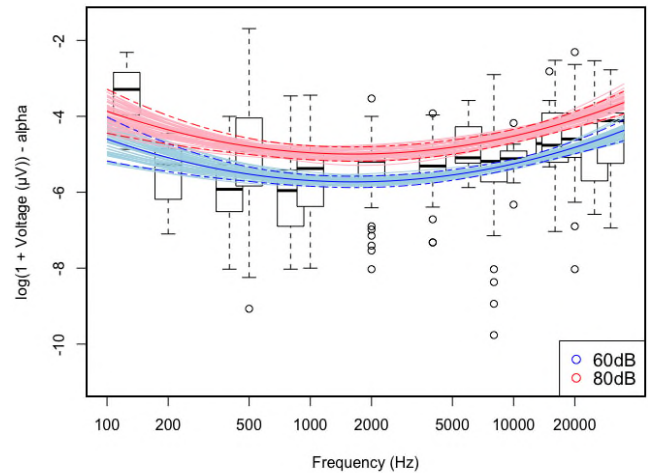

**Figure 4**

Relationship between piezotympanic voltage (fixed-effect corrected data) and acoustic frequency: estimated model for the two sound levels, 60 dB and 80 dB (bold parabolic line) and 50 random models based on the estimated covariance matrix of coefficients (curves in between the two dotted lines and encompassing the bold line).

All the coefficients are highly significant ($p < 0.001$). The standard error of $U_{ijks}$ is 1.109 and the adjustment is measured by the $R^2 = 0.90$ (adjusted 0.89 with 376 degrees of freedom). The F-statistic is 53.57 with a p value smaller than $22.10^{-16}$. The estimated U-shaped curve is represented in fig. 4

The values of the $\alpha_{is}$ fixed effects are distributed between 5.24 and 11.69 with an average value of 8.02 dB. We established a significant ($p < 0.001$) positive correlation between $\alpha_{is}$ and two individual characteristics: "Age" and "Body Mass Index"; There is also a strong correlation between left and right sides.

Other factors could explain these fixed effects, such as pressure on the measurements probe by the ENT practitioner, position of the two electrodes of the probe relative to a unique (or not) collagen bundle (Cf. Si 02), muscular activity of the tensor tympani muscle (Cf. Si 03A) and pressure on the epidermis by the practitioner (Cf. Si 03B).

Taken together, these measurements show that the tympanum responds to acoustic stimulations by isochronous potentials, which we attribute to the tympanum's collagenous fibers. This result corresponds to the outcomes of our measurements on other collagenous fibers such as tendons: knees, Achilles, arm muscles, etc. (Cf. Si 01; Si 14B).

These isochronous tympanic potentials (dubbed "pT") are dependent upon frequency. The voltage decreases from infrasounds to middle frequencies (minimum at 1632 Hz), and then increases along with frequencies for every subset of measures.

Our measures of pT may seem of a too low tension to act on the cochlea; but we must take into account the fact that the result of our measures is diminished by the interposition between the probe and the source of the electric potential of an more or less insulating layer, the epidermis of the eardrum. When we take into account the weakening of conduction by the epidermal tympanic layer, these values seem consistent with our measurements on the eardrum in vivo (hundreds of micro-volts for an acoustical stimulation of approximately 70 or 80 dB SPL).

Furthermore, Harnagea showed in vitro that the response is low or non-existent if the electrodes are arranged in any two points of the biological culture, but this response reached tens of milliVolts if the electrodes are placed at both ends of a single fiber... That was confirmed by Denning et al. (2017

S27 - S6). Thus we can assume that the pT voltages are in the mV range and may reach the DOHC complex (Si09).

We hypothesize that these tympanic voltages created by the piezoelectricity of the collagen fibers of the eardrum are transmitted very rapidly to the OHCs of the cochlea by an electrical pathway. We will now describe this hypothesized "covert path," yet up to now neglected.

## 4. Discussion

### 4.1. The covert path, from tympanum to trickystor

We will show that electrical responses are transmissible via a series of electrical synapses from the tympanum to the apex of the DOHC complex (Deiters Outer Hair Cell complex), where we identify a structure similar to that of a Field Effect Transistor (FET). We dub this structure Trickystor (TkS) due to its complexity.

Gap Junctions are composed of two hemi-channels, each made up of six connexins (Cx). Gap Junctions (GJs)[28] are cytoplasmic conduits possessing large pore size (10–15 Å). They allow communication between the intracellular milieus of two contiguous cells and the passage of small metabolites and signaling molecules (mass $< 2.10^{-27}$ kg) between cells. GJs are very fast conductors able to constitute an electric network. The GJs are especially useful in facilitating electrical transmissions[29]. One of the neuronal functions of GJs is thought to be synchronization between brain cells[30]. The transmission of a signal by means of these electrical synapses is not dependent upon a certain threshold. Further, such transmission is extremely rapid and takes place without diffusion (leakage) into extracellular spaces. It is noteworthy that an electric sinusoidal wave (e.g. pT voltages) can travel along an electrolytic pathway, going through the GJs with minute displacements of ions between adjacent cells. This takes place without global displacements from the first cell to the last and back. Alternating current (AC) voltages cause no net movement into the conductive medium, regardless of its length, since the charge carriers oscillate back and forth in response to an alternating electric field. Nanotube structures might be implicated in the electrical communication by GJs between ear and OHCs[31].

The cell bodies of the osteocytes act as mechanosensors of the petrosal bone. They merge to form a syncytium (based on the Cx43) capable of conveying electrical signals. Electrical transmission between osseous cells always travels in the same direction: from the interior of the bone toward its surface (periosteum)[32]. Electrical signals arising from the piezo-electricity of the tympanum can, thus, be transmitted to the external wall of the cochlea (the spiral ligament, which is a periosteum structure) via the syncytium of the subperiosteal cells.

Through the root cells[33], Cx43 interacts with the Cx26 of the cochlea, thus enabling the transmission of the piezotympanic signal to the cochlear Deiters Cells (DCs). A critical relationship may be established between the mutation of Cx43 proteins and non-syndromic high-frequencies deafness[34].

Yet there are two independent syncytia in the cochlea:
- The connective tissue GJ system of the lateral wall (fibrocytes): The deterioration of this system results in a progressive hypoacousis, especially with respect to high frequency sounds[35]. The Fibroblast Growth Factors (FGFs)[36], which regulate the electrical excitability of cells, appear to have a role in the maintenance of normal auditory function[37].
- The epithelial cell GJ system is composed of several types of supporting cells linked to the *root cells* within the spiral ligament. It is obviously the most important system for transmitting the pT signal. Root cells are present primarily in the basal part of the cochlea, the part devoted to hearing high frequencies. The epithelial cell GJ system is capable of transmitting variations of potential[38] from the root cells to the DCs, and, when it does not function, the OHCs, even if they are normal, lose their effectiveness[39].

---

[6]The longitudinal piezoelectric coefficient for individual fibrils at the nanoscale was found to be roughly an order of magnitude greater than that reported for macroscopic measurements of tendon, the low response of which stems probably from the presence of oppositely oriented fibrils, as confirmed here.

Thus, active cochlear amplification is dependent on the GJs of supporting cells[40] - [41]: Genetic[42] or experimental alterations of either the structure of root cells or of several connexins [Cx26 (GJB2), Cx30 (GJB6), Cx31, Cx32, Cx43] have been shown to result in non-syndromic deafness[43 - 44 - 45]. Purely metabolic explanation of their usefulness seems insufficient to explain why this is so. The number of Cx26 and Cx30 declines from the cochlear apex to its base, but this finding does not weaken the hypothesis that these GJs play an essential role for all frequencies: Either mutations or a blocking[46] of Cx26 produces a reduced, or absent, distortion product of otoacoustic emission (DPOAE) and hearing loss at all frequencies.

That apparent discordance should be clarified by their role in the functioning of the Trickystor (below).

## 4.2. The Trickystor

Travelling waves cause the stereocilia crowning the OHCs to move, producing a mechano-electrical transduction. The resulting electrical signals, if their frequency is under a few kHz, pass through the cuticular bilayer.

Our experiments have demonstrated the piezoelectricity of the eardrum and its adjacent bony structures. We have also presented a probable route of transmission of the tympanic electrical signals (pT) via electrical synapses (GJs) up to the DOHC complex, It has been proposed that cochlear support cells interact with hair cells in a manner similar to interneurons or astrocyte interactions with neurons in the central nervous system[47] - [7]. Each OHC is surrounded by five DCs (fig. 5). Its base is supported by the cupular body of a Deiters Cell ($DC_5$) and its ciliated apex is bordered by four phalangeal apexes from four other $DC_{is (i=1..4)}$: on the right ($DC_1$), inside ($DC_2$), on the left ($DC_3$), outside ($DC_4$); each being different from the $DC_5$.

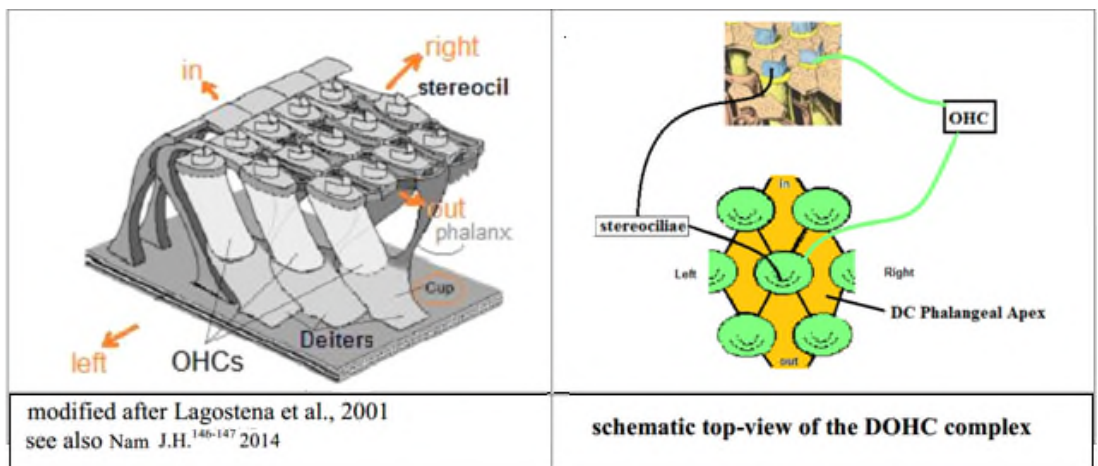

Figure 5
3D and top view of the DOHC-complex organization

Every component of an Organic Field Effect Transistor (OFET) is present in the apex of the DOHC complex (Fig. 6).

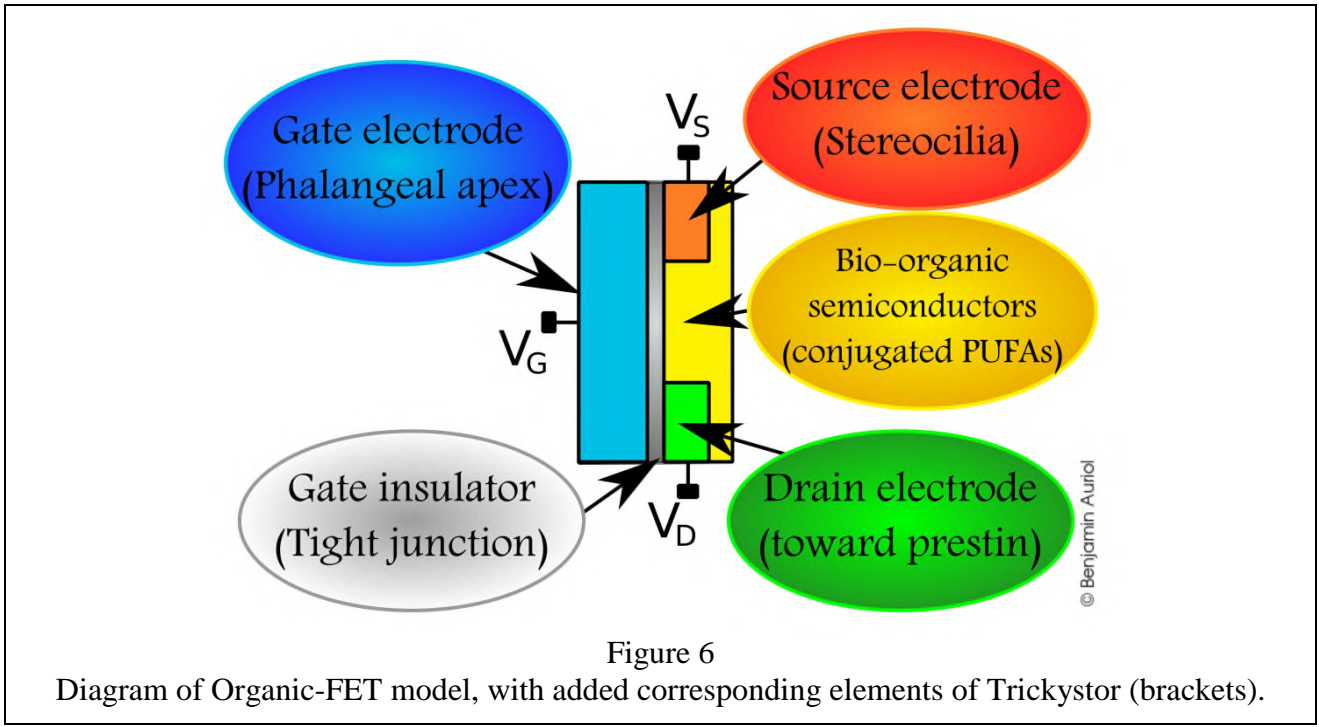

Figure 6
Diagram of Organic-FET model, with added corresponding elements of Trickystor (brackets).

We will consider its elements in turn: the source (Cf. Si 10 A), the semiconductor channel, the gate with its electrical insulation from the overall structure, and, finally, the drain.

The source is identical to what has been shown in the classical literature: Electrical potentials result from stereociliae movement (flexoelectricity) and from the biasing potential.

A common feature of Organic FET materials is the inclusion of a conjugated π-electron system. This system serves as the active semiconducting layer and facilitates the delocalization of orbital wave functions. In its original meaning, a conjugated system is a molecular entity whose structure may be represented as a system of alternating single and multiple bonds. In such systems, conjugation is the interaction of one π-orbital with another across an intervening σ-bond. Homoconjugation is defined as "an orbital overlap of two π-systems separated by a non-conjugating group, such as CH2" [48- 49]. Conjugation and homo-conjugation alike give semiconductor properties to a biological molecule. When biological poly-unsaturated fatty acids (PUFAs) are either conjugated (R-CH=CH-CH=CH-R') or homoconjugated (R-CH=CH-CH2-CH=CH-R), they have the properties of a semiconductor[50].

The cuticular bilayer (fig.07) encompasses semi-conductors such as phospholipids, conjugated linoleic acid, conjugated linolenic acids, or docosahexaenoic acid ethyl ester-d5 as well as other conjugated and homoconjugated PUFAs. There is an inverse association between hearing loss and higher intakes of long-chain n−3 PUFAs and regular weekly consumption of fish[51]. Modifications of the PUFAs by genetic mutations, for example, "peroxisome biogenesis disorders " or "X-linked adrenoleukodystrophy", have deleterious consequences on the auditory processing of high frequencies (Cf. Si  10D).

Fig. 07 presents two conductive pathways capable of passing charge carriers through the cuticular membrane: The ionic channels are relevant for frequencies below 3 kHz[52] but ineffective for higher frequencies; the semiconductor channels are most likely to intervene for high frequencies, up to more than 200 kHz.

In fact, these two connecting structures might be topographically associated since conjugated PUFAs are close neighbors of ionic channels. PUFAs are incorporated into the lipid bilayer near to, but not included within, the pore domain. They affect voltage transition electrostatically. If the charge is switched, these electrostatic interactions accomplish opposite effects  (Cf. Si 10B  and 10C).

The leaflets of the bilayer are primarily composed of phospholipids, and anti-phospholipids can negatively affect hearing (Cf. Si 10D). Chlorpromazine, which intercalates into the inner leaflet of the phospholipid bilayers, alters OHCs electro-motility without a known direct action on prestin; According to Ricci et al.[53] "The conductance of the Mechano Electrical Transducer channels changes along the tonotopic position within the cochlea, suggesting differential requirements at different frequencies". Thus, the intervention of the phospholipids concerns specifically neither the action of the stereocilia, nor that of the prestin, but rather that of the cuticular bilayer.

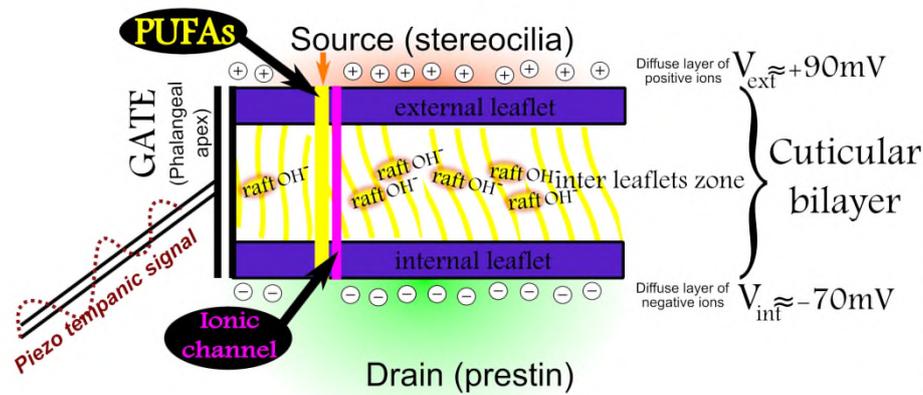

Figure 07
Simplified diagram of the cuticular bilayer

The doping of a semiconductor material consists of introducing into its matrix very small quantities of other material comprising different charge carriers. Small numbers of them can change the ability of a semiconductor to conduct electricity. In the case of conjugated PUFAs, the doping material may consist of a system of anionic cholesterol rafts nested between the two leaflets of the bilayer[54]. These cholesterol rafts modify the passage of electrons through the bilayer in certain directions and also the voltage dependence of the prestin (Cf. Si 10C).

The lipid bilayer is associated with an electronic double layer: endocochlear (+ 90 mV) versus intracellular (-70 mV) potential; It has a non-linear capacitance dependent upon the voltage applied. That is an essential component of a beneficial **biasing** imposed on the TkS. The (Cx26$^{+/-}$/ Cx30$^{+/-}$) digenic mutation, which decreases that bias, results in high frequencies hearing loss[55].

The gate is represented by Deiters phalanxes with their high microtubules content. These microtubules cause negative differential resistance, improve electric connectivity between their two ends, and amplify the critical frequency of the transferred signals[56].

In physiological conditions, there is a strict insulation, chemical as well as electrical, between phalangeal apexes of 4 DCs and the cuticle of an embedded OHC (fig. 5). This being the case, no electric current will flow from one to the other of their apical membranes.
However, this border, the tight junction (TJ) and other elements of the Apical Junctional Complex cannot prevent a hydrophobic intercellular electrostatic coupling unrelated to any GJ. Thus, the stereociliae signal can be amplified by the isochronous piezotympanic (pT) signal acting on semiconductors of the cuticular membrane. TJs between OHCs and DCs are, indeed, critical for normal functioning of the organ of Corti; mutations of the TJP2 gene cause autosomal dominant non-syndromic hearing loss[57] (Cf. Si 10F; Cf. also Si 10G).

As in the classical FET schema, the electric signal from the source must reach the drain (prestin) (Cf. Si 10H) within the lateral wall of the cell and stimulate it. In our view, when passing through the cuticular bilayer, the signal is driven by the TkS. Voltage variations due to the stereociliae alternatively shorten and lengthen the prestin located in the latero-basal wall of the OHCs. In the case of the highest frequencies, we make the assumption that the signal is amplified by the pT coming from the external ear (eardrum and bone collagen). The problem of the presence of prestin in the vestibular system dedicated to some infrasounds is examined in Si 10I.

## 4.3. Overview of the "covert path"

The TkS is probably a decisive element for refreshing the acoustic signal (especially for high frequencies, or very low ones), for enhancing cochlear amplification, and for frequency analysis (tuning) of the audio signal.

We propose (Fig. 08) the design of an equivalent circuit which presents the classical pathway (in black) supplemented by the "covert path" (in red).

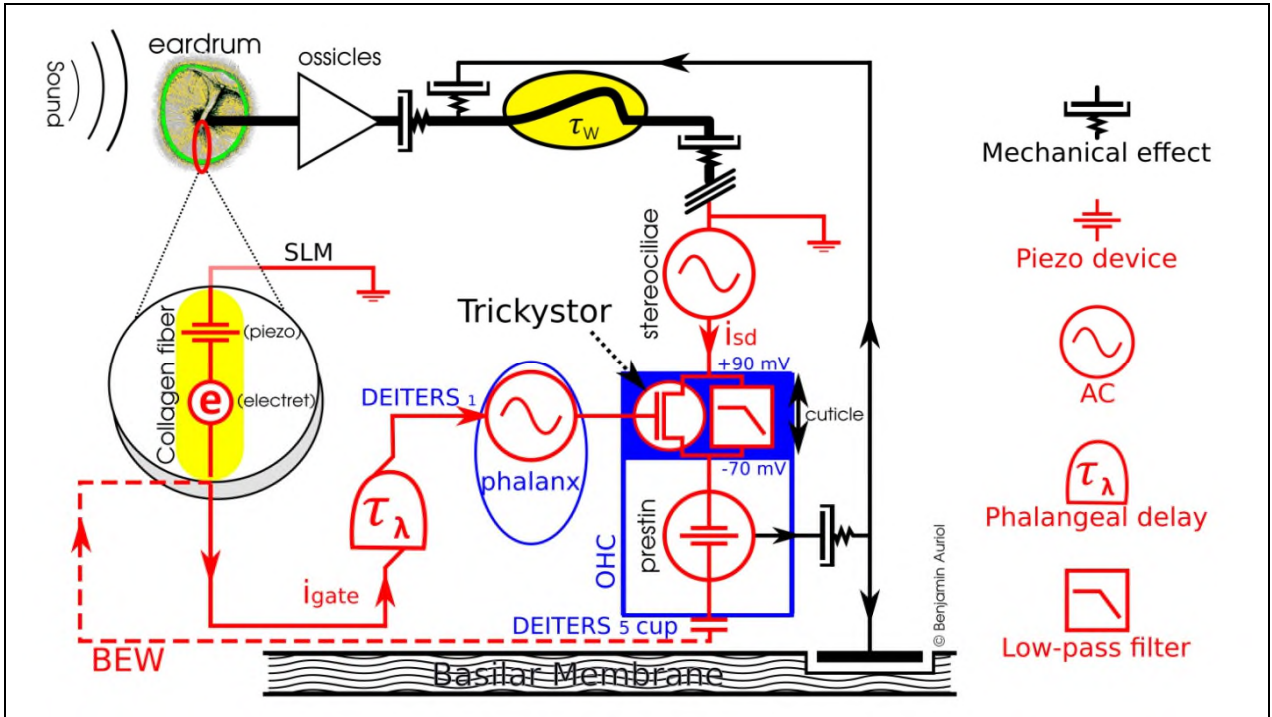

Figure 08

Equivalent circuit: classical pathway (in black); "covert path", our model (in red).
 SLM (Superior Ligament of Malleus): Collagenic ligament that crosses from the head of the malleus to the roof of the tympanic cavity (tegmentum attici).
*Igate* : Input Gate toward phalangeal cuticle; *Isd* : Input into OHC of stereocilia flexoelectric signal.
BEW: hypothetic Backward Electrical Wave (Cf. Si 12 OAE vs BEW).

## 4.4. '*Sine Qua Non*': demonstration arguments for the covert path

Every step of the covert path is indispensable for good hearing, and any deficiency (either genetic, experimental or toxic) in any of the steps of the covert path produces hearing impairment, mainly in the high frequencies.

In the following figure (Fig. 09), we number the locations of consecutive steps whose modification has a hearing loss effect. For each of these steps, we then listed a list of important publications linking these hearing losses to these alterations.

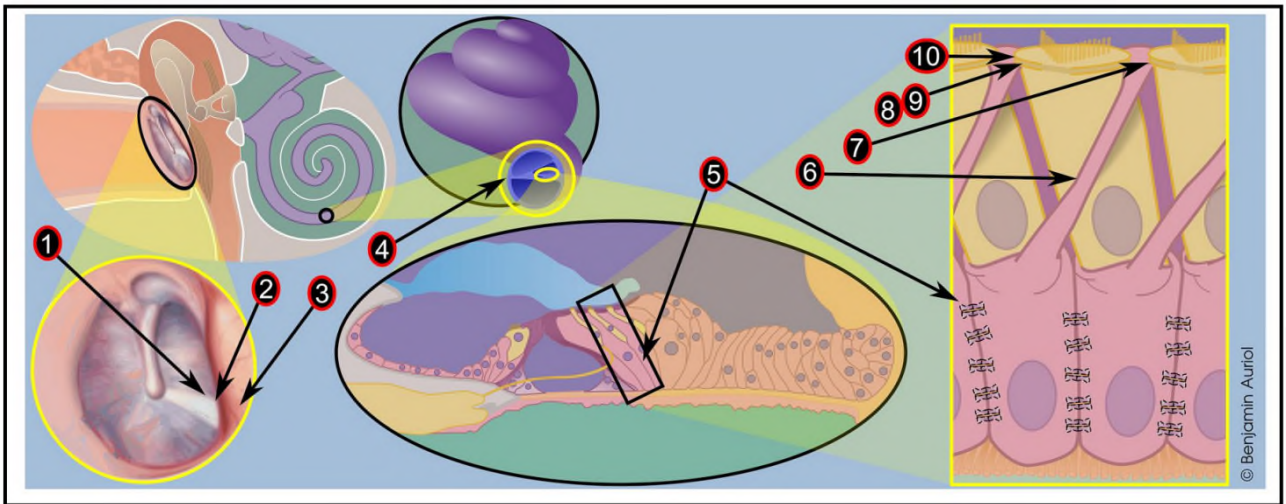

| Step (1) | Step (2) | Step (3) | Step (4) | Step (5) | Step (6) | Step (7) | Step (8) | Step (9) | Step (10) |
|---|---|---|---|---|---|---|---|---|---|
| Collagen II | Annulus T. | GJ Cx43 | Root cell | GJs Cx26,30 | Phalanx | TJs | Semi-conductors | Doping | Biasing |

Figure 09 (schéma)

*'Sine Qua Non'* demonstration of the covert path

For each step indexed in fig. 09, we now give a brief description, and then we cite the references establishing the negative effect of any local flaw on auditory functioning:

- *Step 1 Collagen II* of the eardrum and mastoid. Any flaw in the structure of type II collagen, accompanied by a defect in its piezoelectricity, causes hearing loss (Liberfarb RM, et al., The Stickler syndrome *Genet. Med*,.**5,** 21-27 (2003) and *Omim120140*)

- *Step 2 The annulus fibrosus tympanicus* is the thickened peripheral rim of the pars tensa of the tympanic membrane; It is connected with the bony sulcus tympanicus via radial fibers bundles, which continue directly into the tympanic bone. Osteoma of the osseous and fibrous annulus tympanicus can cause hearing impairment (Uno Yoshihumi, The attachment structure of the guinea pig tympanic membrane, Auris Nasus Larynx, 27,45-50(2000) and He Z., Vibration measurements on the widely exposed gerbil eardrum, Biomedical Engineering, partial fulfilment of Master of Engineering, McGill University (2012)).

- *Step 3 Mutations in GJA1 (connexin 43)* are associated with non-syndromic autosomal recessive deafness ( Liu, X. Z., et al., Mutations in GJA1 (cx43) are associated with non-syndromic autosomal recessive deafness. Hum Molec Genet 10: 2945-2951, 2001).

- *Step 4 Crossing spiral ligament* via **Root-cells** may be central to pathological processes associated with various forms of hearing loss (Jagger DJ, Forge A. The enigmatic root cell - emerging roles contributing to fluid homeostasis within the cochlear outer sulcus. *Hear Res.* **303,** 1-11 http:\www.sciencedirect.com\science\article\pii\S0378595512002523 2013) .

- *Step 5 GJs* (Cx26, Cx30) Mutations in either Cx26 or Cx30 are the major cause of non-syndromic prelingual deafness in humans.

The active cochlear amplification is dependent on the connexin 26 : Hong Bo Zhao et al.[8] have shown experimentally that active cochlear amplification is very markedly reduced when Cx26, and the GJs which use it, disappear or are deteriorated. DPOAE and OHC electromotility were reduced, a hearing loss initiated at high frequencies and then extended to the other frequencies (Zong L, Chen J, ZhuY, Zhao HB, Progressive age-dependence and frequency difference in the effect of gap junctions on active cochlear amplification and hearing, Communications, 489, 223-227 (2017), Zhu, Y. et al., Active cochlear amplification is dependent on supporting cell gap junctions, *Nat Commun*,**4,** 1786 (2013)**,** Omim 604418 and Chang Q, Tang W, Ahmad S, Zhou B, Lin X, Gap Junction Mediated Intercellular Metabolite Transfer in the Cochlea is Compromised in Connexin30 Null Mice,. *PLoS ONE* **3 (12)**, e4088 (2008)).

- *Step 6 Phalanx of Deiters* (GATE). Destruction of the phalanx cytoskeleton annihilates the electric effect of the DCs on the OHCs (Yu N, Zhao HB, Modulation of Outer Hair Cell Electromotility by Cochlear Supporting Cells and Gap Junctions. PLoS ONE 4(11): e7923 (2009))**.**

- *Step 7 Tight Junctions* between phalanx of Deiters Cell and OHC cuticle; Mutation of the *TJP2* gene causes autosomal dominant non-syndromic hearing loss (ADNSHL) (Kim MA, et al. Genetic Analysis of Genes Related to Tight Junction Function in the Korean Population with Non-Syndromic Hearing Loss. PLoS ONE 9(4), e95646 (2014). Wilcox ER et al., Mutations in the Gene Encoding Tight Junction Claudin-14 Cause Autosomal Recessive Deafness DFNB29, Cell, 104: 165-172 (2001).Tang VW. Proteomic and bioinformatic analysis of epithelial tight junction reveals an unexpected cluster of synaptic molecules. *Biology Direct.* 2006;1:37**.** Itajiri S, Katsuno T. Tricellular Tight Junctions in the Inner Ear. *BioMed Research International.* 2016;2016:6137541)**.**

---

 Y Zhu, J. Chen, C. Liang, L. Zong, J. Chen, R.O. Jones and H.B. Zhao, Connexin 26 [GJB2] deficiency reduces active cochlear amplification leading to late-onset hearing loss, Neuroscience 284, 719-729 (2015)

- *Step 8 Semi-conductors* (Conjugated PUFAs and phospholipids, etc.); Genetic, dietary or toxic deficiencies of conjugated PUFAs in cuticular phospholipids, induce a downward sloping audiometric pattern (OMIM# 253260 and 609019., Wolf B, Spencer R, Gleason T., Hearing loss is a common feature of symptomatic children with profound biotinidase deficiency. J Pediatr 2002, 140,2:242–246**,** Van Veldhoven PP, Biochemistry and genetics of inherited disorders of peroxisomal fatty acid metabolism, J. of Lipid Research, Thematic Review Series: Genetics of Human Lipid Diseases, 51, 2010 , p. 2885, Wanders RJA, Peroxisomes, lipid metabolism, and peroxisomal disorders, ASHG 2004 Meeting Toronto, Molecular Genetics and Metabolism, 83, 1–2, September–October 2004: 16–27**,** Braverman NE et al., Peroxisome biogenesis disorders in the Zellweger spectrum: An overview of current diagnosis, clinical manifestations, and treatment guidelines, Mol Genet Metab 117 : 313-21 (2016), Zempleni J, Hassan YI, Wijeratne SS, Biotin and biotinidase deficiency, Expert Rev Endocrinol Metab **3** : 715–24 (2008), March J, Advanced Organic Chemistry reactions, mechanisms and structure (3rd ed.). New York: John Wiley and Sons (1985), Hush, N. S., An Overview of the First Half-Century of Molecular Electronics. Annals of the New York Academy of Sciences, 1006: 1–20 (2006), Inzelt, György "Chapter 1: Introduction". In Scholz, F. Conducting Polymers: A New Era in Electrochemistry. Monographs in Electrochemistry. Springer : 1–6. ISBN 978-3-540-75929-4. (2008), Bard Allen J., Inzelt György, Scholz Fritz, Electrochemical Dictionary : cuticular phospholipids, Springer Science and Business Media. (2008), Engelman DM. 2005. Membranes are more mosaic than fluid. Nature 438: 578–580, Jacobson K, Mouritsen OG, Anderson RGW. 2007. Lipid rafts: At a crossroad between cell biology and physics. Nat Cell Biol 9: 7–14. and Coskun U., Simons K. 2010. Membrane rafting: From apical sorting to phase segregation. FEBS Lett 584).

- *Step 9 Doping* *by cholesterol rafts*. The modulation of Voltage-gated calcium channels (VGCCs) and Big Potassium channels (BK) currents by cholesterol, and the associated changes in hair cell excitability may have implications for sensorineural hearing loss.(Purcell EK, Liu L, Thomas PV, Duncan RK. Cholesterol Influences Voltage-Gated Calcium Channel and BK-Type Potassium Channels in Auditory Hair Cells. Dryer SE, ed. *PLoS ONE;* 6(10):e26289. doi:10.1371/journal.pone.0026289 (2011), https://en.wikipedia.org/wiki/Niemann%E2%80%93Pick_disease,_type_C.**,** Oghalai JS., Pereira FA., and Brownell WE., Tuning of the Outer Hair Cell Motor by Membrane Cholesterol, J Biol Chem., 282(50): 36659–36670. (2007)[9]).

- *Step 10 Biasing*. The lipid bilayer is associated with an electronic double layer (endocochlear potential), which is an essential component of a beneficial biasing imposed on the TkS. The ($Cx26^{+/-}$/ $Cx30^{+/-}$) digenic mutation, which decreases that bias, results in high frequencies hearing loss (Mei L et al., A deafness mechanism of digenic Cx26 (GJB2) and Cx30 (GJB6) mutations: Reduction of endocochlear potential by impairment of heterogeneous gap junctional function in the cochlear lateral wall, *Neurobiol Dis.* **108**, 195-203 (2017)).

Thus the preceding analysis gives rise to a *Reduction ad absurdum* of our theory of covert path. In fact, if the covert path does not exist as a necessary complement to the TW of Von Bekesy, the dysfunctioning of any of the steps of the chain that we describe and which our theory assumes would have practically no effect, since the chain does not exist its stages should not exist either. The dysfunctioning of any of the steps listed above should not affect hearing. The analysis of the 10 "steps" above, supported notably by the works of genetics, proves the opposite.

# 5. Conclusion and perspectives

The electric signal generated by tympanic collagen fibers is not conceived as the alternative origin of the mechano-sensation in the auditory system, but rather as a significant electronic contribution (Cf. Si 10E , Si 13, Si 19 and Si 20).

Our experiments have shown that the tympanum has piezoelectric properties that engender an electrical signal in response to acoustic vibrations. This signal, which is frequency-dependent, is, then, carried to the outer wall of the cochlea and from there to the DCs by means of electrical synapses (various GJs and their connexins).

The piezoelectricity of the tympanum opens up the perspective of an electrical synergistic pathway of sound transmission heretofore unknown (the covert path). This pathway from the tympanum to the cochlea is capable of contributing significantly to hearing, especially to hearing the highest frequencies, as it has a determining effect on the amplification and tuning attributed to the basal OHCs (Cf. Si 11B).

Our hypothesis of an electric pathway does not negate the established theory of sound transmission but rather expands it. For it is our idea that the mechanical and the electrical transmission of sound work together[58] - [59] to produce optimal hearing (especially for high frequencies). Thus, our findings pave the way for a better understanding of the hearing process and have important implications for both theory and practice. It is well known that age-related hearing loss primarily involves high frequency sounds, the very sounds we believe are mainly transmitted by the electrical pathway. Our work opens perspectives in the understanding of hearing and in the treatment of hearing problems.

---

[9] doi: 10.1074/jbc.M705078200; Cf. also Si 2.

So the discovery of this electric transmission of sound may elucidate certain as yet unexplained phenomena of auditory physiology. For example, it may lay the groundwork for a better understanding of otoacoustic emissions (OAEs). No satisfactory explanation has yet been found for the backward propagation of OAEs (elusive backward travelling wave) [60 - 61]. Our theory may also shed light on hyperacusis of children, which is associated with larger amplitude OAEs but with no other auditory factors[62]. Furthermore despite ossicular blockage by the mesenchyme until after birth, it has been shown (see Si 08B) that the foetus hears and memorizes the sounds of its environment several weeks before birth[10]! This makes it difficult to understand the existence - though well demonstrated - of foetal hearing from the $22^{th}$ amenorrhea week[11 - 12]. So Hill concludes that "any prenatal conduction to the cochlea must be mediated through bone conduction". However, the mechanism of bone conduction is itself poorly understood, so that the explanation of foetal hearing by bone conduction would simply move the problem on. It therefore seems that another conceivable mechanism would be the "covert path".

In fact, it is likely that knowledge of this new pathway can shed light on how sea mammals and bats use very low or very high frequencies for echolocation. In the case of cetaceans, hearing is dependent upon a "collagenous-fatty acoustic pathway" which, according to our theory, primarily uses the piezo-electricity of collagen to amplify frequencies up to more than 250 kHz (Cf. Si 08). The eardrum of terrestrial mammalians and the collagenous-fatty acoustic bodies of sea mammalians share a double function: on the one hand, a mechanical mobilization for medium frequencies, and, on the other, a piezoelectric behavior of collagen fibers for the perception of very high and very low frequencies.

This new findings might have an important impact on the future **development of hearing aids**. Such a practical application would be very useful for an ageing population.

_______________________________________________________________________________________

## Supplementary Information

This article is very shortened and limited to what is most recent in our theory and results. So we propose some supplementary information (Si) to expand and discuss several aspects of our experimental and theoretical work.

The authors declare no competing financial interests.
This work has not been published previously and is not under consideration for publication elsewhere.

All authors have approved the manuscript in its present form. Supplementary Information are available on /Supplementary Information
 and Source Data are available on 10.6084/m9.figshare.5671807.
Two figures summarizing the main results of this paper are included as Fig 04 and Fig 08. Readers are welcome to comment online.

**Author Contributions**

BMA Proposed the idea of the "covert path" (with piezotympanic source), contributed to the physiological theoretical work;

JB, J-MB, DD and CT designed the experiments;
BMA, JB, J-MB, DD, PP and BL performed measurements;
JB, J-MB, JPF, and FG analyzed data;
BA made the artistic figures;
BMA, JB, JPF, LD, BB, FG, EJ wrote the paper;
RR did the acoustic instrumental calibration.
The authors declare no competing financial interests.

## Acknowledgments

We are very thankful to the volunteers which gave their time for participating to the experimental measurements and to the complementary tests.

We thank for scientific assistance : INP of CNRS; Andrews S., PhD, LSU Medical School;  Auriol J.B., PhD, BD (New Jersey, US); Azaïs C., PhD, ex-LAMI, Toulouse University; Banquet J.P., MD, PhD, CNRS UMR, Paris; Bibé B., Emeritus Research Director, INRA, genetics and wording; Calvas F., MD, CIC, Inserm, CHU-Purpan Toulouse; Chernomordik L.V., Ph.D., Senior Biomedical Researcher, NICHD, NIH, Bethesda, USA; Csillag P., ENSEEIHT - Toulouse France; Demmou H, Senior Researcher, LAAS – CNRS, Toulouse, France; Faurie-Grepon A., MD, CIC, Inserm, CHU-Purpan Toulouse, France; Gaye-Palettes J.F., MD, EHPAD Isatis, Quint-Fonsegrives; Harnagea C., PhD, INRS, Montreal, Canada; Lagrange D., Electronic engineer, LAAS, Toulouse; Legros C., Emeritus Pr –Jean Jaurès University, Toulouse; Marlin S., INSERM, Centre de Réf. des Surdités Génétiques, CHU Necker, Paris; Marmel F., PhD, INCyL, Salamanca, Spain; Merchan M.A., PhD, INCyL, Salamanca, Spain; Millot M., PhD, LNCMI, UPS Toulouse; Mitov M., PhD, Director of Research (CNRS), leader of the liquid-crystal group (CEMES, Toulouse) ; Morell M., PhD, INM, Inserm Unit 1051; Norman B., Distinguished Pr Emeritus, University of South Carolina, Columbia, USA; Pr Petit C., for her outstanding teaching and contributions  to our field; Picaud F., PhD, MCF., University of Franche-Comté, UFR des Sciences et Techniques ;  Portalès P., Electronic engineer,  Airbus Industry Toulouse, France; Szarama K.B., PhD, NIH, NIDCD, Bethesda, USA; Weeger N., ex-Pdt of Wikimedia-France.; Guijo-Pérez B. English Translator.

We thank for practical assistance:
The British Library - London; Le Neillon M., Connectic System - Toulouse; Leroy L., Veterinary Nurse – Tournefeuille, France; Auriol family for their encouragements, implication and advices (Nanou, auteure; Emmanuelle Auriol, PhD, Pr TSE).